# Effect of multinary substitution on electronic and transport properties of TiCoSb based half-Heusler alloys

Mukesh K. Choudhary[1, 2] and P Ravindran[1, 2, 3, 4, *]

[1]Department of Physics, Central University of Tamil Nadu, Thiruvarur-610005
[2]Simulation Center for Atomic and Nanoscale Materials, Central University of Tamil Nadu, Thiruvarur-610005
[3]Department of Materials Science, Central University of Tamil Nadu, Thiruvarur-610005
[4]Department of Chemistry, Center for Materials Science and Nanotechnology, University of Oslo, P.O. Box 1033 Blindern, N 1035 Oslo, Norway
[*]email: raviphy@cutn.ac.in

**Abstract.** The electronic structures of $Ti_xZr_{x/2}CoPb_xTe_x$, $Ti_xZr_{x/2}Hf_{x/2}CoPb_xTe_x$ ($x = 0.5$), and the parent compound TiCoSb were investigated using the full potential linearized augmented plane wave method. The thermoelectric transport properties of these alloys are calculated on the basis of semi-classical Boltzmann transport theory. From the band structure calculations we show that the substitution of Zr,Hf in the Ti site and Pb and Te in the Sb site lower the band gap value and also change the indirect band (IB) gap of TiCoSb to the direct band (DB) gap. The calculated band gap of TiCoSb, $Ti_xZr_{x/2}CoPb_xTe_x$, and $Ti_xZr_{x/2}Hf_{x/2}CoPb_xTe_x$ are 1.04 eV (IB), 0.92 eV (DB), and 0.93 eV (DB), respectively. All these alloys follow the empirical rule of 18 valence-electron content which is essential for bringing semiconductivity in half Heusler alloys. It is shown that the substitution of Hf at the Ti site improve the *ZT* value (~1.05) at room temperature, whereas there is no significant difference in *ZT* is found at higher temperature. Based on the calculated thermoelectric transport properties, we conclude that the appropriate concentration of Hf substitution can further improve the thermoelectric performance of $Ti_xZr_{x/2}Hf_{x/2}CoPb_xTe_x$.

## INTRODUCTION

Rapidly increasing population and the relatively increase of energy demand has motivated the scientific community to develop the alternative technique to convert the waste form of energy into the useful clean form of energy. Half-Heusler (HH) alloys are considered as one of the potential candidates for the waste heat recovery.[1] HH alloys has general formula of *XYZ* in which *X* and *Y* typically are transition metals and *Z* is a main group element. Semiconducting behavior in the HH alloys is arising due to the transfer of valence electron from the less electropositive element *X* to more electropositive element *Y* and *Z* which gives the close shell configuration i.e. a $d^{10}$ for *Y* and $s^2p^6$ configuration for *Z*. The efficiency of the thermoelectric material can be calculated by the dimensionless figure of merit $ZT= S^2\sigma T/\kappa$, where S is Seebeck coefficient, σ is electrical conductivity, $\kappa$ is total thermal conductivity i.e. electronic and thermal conductivity ($\kappa_e + \kappa_l$).[2, 3] Extensive research has been done on the TiCoSb based half –Heusler alloy due to its large value of Seebeck coefficient and good electrical conductivity. However this alloy is having large thermal conductivity which restrict its use in thermoelectric applications. There has been lot of efforts made to reduce the thermal conductivity in this material and in this work we have focused on the isoelectronic substitution of element in the Ti site of the parent compound TiCoSb. Further, the Sb atoms were substituted with Pb and Te by preserving 18 VEC. Substitution of atoms with dissimilar atoms preserving 18 VEC create extra phonon scattering center into the lattice due to atomic size mismatch and hence one can expect reduction in the thermal conductivity of the materials.[4,5]

# COMPUTATIONAL DETAILS

We have performed the structural optimization using Vienna *ab-initio* simulation package (VASP)[6,7] within the projector augmented plane wave (PAW) method to find the ground state structure. The Perdew-Burke-Ernzerhof generalized gradient approximation (GGA)[8] is used for the exchange correlation potential. The energy convergence criterion was chosen to be $10^{-5}$ eV and the cut off energy of the plane wave was set to be 400 eV, and the pressure on the cell had minimized within the constraint of constant volume. A supercell approach was adopted for the multinary substitution in the parent compound TiCoSb. A 12x12x12 **k**-mesh was used for the structural optimization. Full-potential augmented plane wave method as implemented in the Wien2k code[9] was used to find the accurate electronic structure for transport properties calculations with the very large **k**- mesh of 24x24x24. The thermoelectric transport properties such as Seebeck coefficient, electrical conductivity, thermal conductivity, and thermoelectric figure of merit were calculated using the BoltzTraP[10] code using the output from Wien2k calculations.

**TABLE 1.** Optimized lattice constants (**a**), band gap values ($E_g$), and thermoelectric figure of merit (ZT) of TiCoSb, $Ti_{0.5}Zr_{0.5}CoPb_{0.5}Te_{0.5}$ and $Ti_{0.5}Zr_{0.25}Hf_{0.25}CoPb_{0.5}Te_{0.5}$. The experimental value of lattice parameter for TiCoSb is given in bracket.

| HH alloys | a(Å) | $E_g$ (eV) | ZT 300 | ZT 700 |
|---|---|---|---|---|
| TiCoSb | 5.87 (5.90) | 1.04 (IB) | 0.98 | 0.95 |
| $Ti_{0.5}Zr_{0.25}Hf_{0.25}CoPb_{0.5}Te_{0.5}$ | 6.03 | 0.93 (DB) | 1.05 | 0.97 |
| $Ti_{0.5}Zr_{0.5}CoPb_{0.5}Te_{0.}$ | 6.04 | 0.92 (DB) | 0.98 | 0.96 |

# RESULTS AND DISCUSSION

Figure 1(a) show the conventional cell of the TiCoSb, $Ti_xZr_{x/2}CoPb_xTe_x$ and $Ti_xZr_{x/2}Hf_{x/2}CoPb_xTe_x$ ($x = 0.5$). These compounds crystallize in the cubic structure with the space group $F\bar{4}3m$ (No.216). The ground state structural parameters were calculated by relaxation of both lattice parameter and atomic positions (see the Fig. 1(b)). The optimized equilibrium lattice parameters, calculated band gap values for pure and multinary substituted TiCoSb are summarized in Table 1.

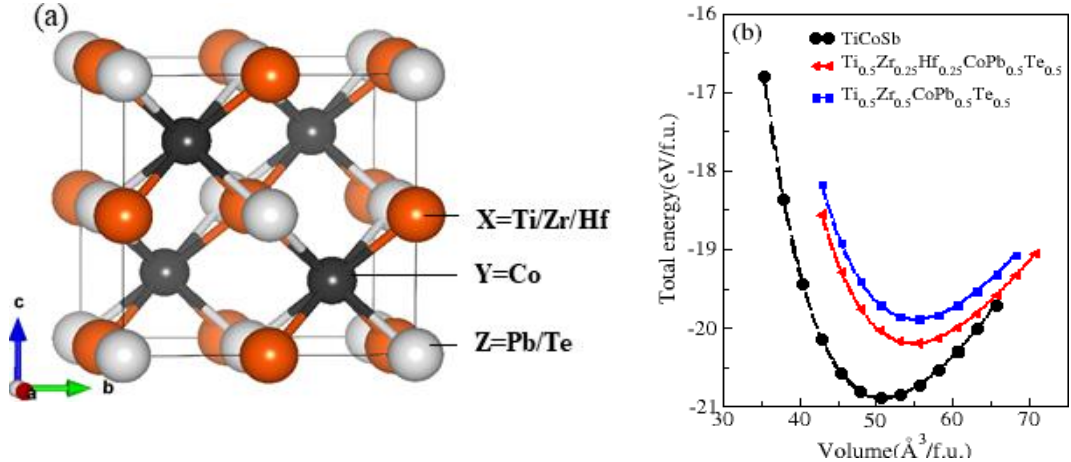

**FIGURE 1.** Crystal structure and unit cell volume vs total energy curve for TiCoSb, $Ti_{0.5}Zr_{0.25}Hf_{0.25}CoPb_{0.5}Te_{0.5}$ and $Ti_{0.5}Zr_{0.5}CoPb_{0.5}Te_{0.5}$

The unit cell of HH alloy *XYZ* contains four formula units with *X* (Ti, Zr, Hf) *Y* (Co) *Z* (Pb, Te) atoms located at the 4a: (0, 0, 0), 4d: (3/4, 3/4, 3/4), and 4c: (1/4, 1/4, 1/4), positions respectively. The calculated equilibrium lattice constants for TiCoSb, Ti$_x$Zr$_{x/2}$CoPb$_x$Te$_x$ and Ti$_x$Zr$_{x/2}$Hf$_{x/2}$CoPb$_x$Te$_x$ are $a$ = 5.871, 6.04, and 6.03Å, respectively. Electronic structure calculations have been performed to understand the electronic ground state and nature of bandgap of these systems. Figures 2(a), 2(b) and 2(c) show the calculated electronic band structure for TiCoSb, Ti$_{0.5}$Zr$_{0.25}$Hf$_{0.25}$CoPb$_{0.5}$Te$_{0.5}$ and Ti$_{0.5}$Zr$_{0.5}$CoPb$_{0.5}$Te$_{0.5}$, respectively closer to their band edges i.e. from -1.5 eV to 1.5 eV. The band structure is plotted along certain high symmetry directions in the irreducible Brillouin zone (IBZ). It can be noticed that the parent compounds TiCoSb show indirect band gap behavior with valence band maximum (VBM) at Γ and the conduction band minimum (CBM) at X point of IBZ of FCC lattice. In contrast, the electronic structure of TiCoSb with Zr, Hf substitution at Ti site and Pb, Te substitution at Sb site show the direct band feature with both VBM and CBM are at Γ point in the IBZ of simple cubic and tetragonal lattice for Ti$_{0.5}$Zr$_{0.25}$Hf$_{0.25}$CoPb$_{0.5}$Te$_{0.5}$ and Ti$_{0.5}$Zr$_{0.5}$CoPb$_{0.5}$Te$_{0.5}$. The corresponding band gap values estimated from our electronic band structures of Ti$_{0.5}$Zr$_{0.25}$Hf$_{0.25}$CoPb$_{0.5}$Te$_{0.5}$ and Ti$_{0.5}$Zr$_{0.5}$CoPb$_{0.5}$Te$_{0.5}$ are 0.93 eV (DB) and 0.92 eV (DB), respectively.

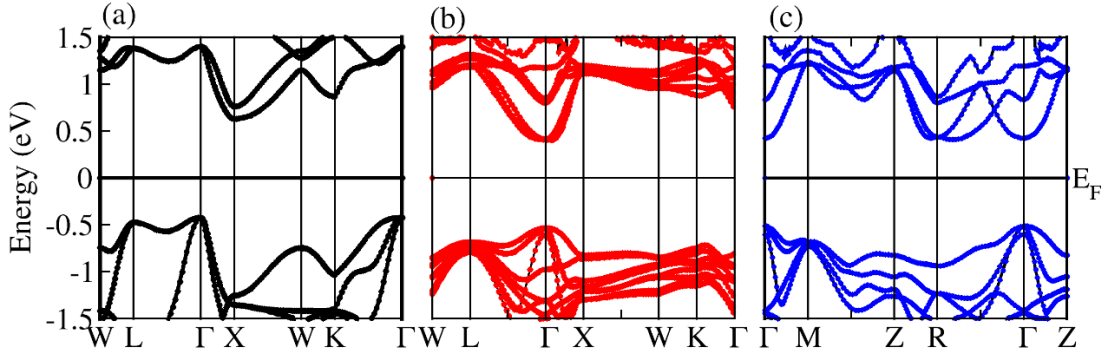

**FIGURE 2**. Band structure of (a) TiCoSb (b) Ti$_{0.5}$Zr$_{0.25}$Hf$_{0.25}$CoPb$_{0.5}$Te$_{0.5}$ (c) Ti$_{0.5}$Zr$_{0.5}$CoPb$_{0.5}$Te$_{0.5}$

Our calculated band gap value for TiCoSb is 1.04 eV which is in good agreement with previous report.[11] It may be noted that the multinary substitution can also bring the semiconducting behavior in HH alloys if it follow the 18 VEC rule. In order to study the thermoelectric properties of HH alloys we have used our calculated electronic structures. We have studied the transport properties of these alloys with respect to chemical potential as well as with temperature to understand the role of electron/ hole doping and temperature effect on transport properties.

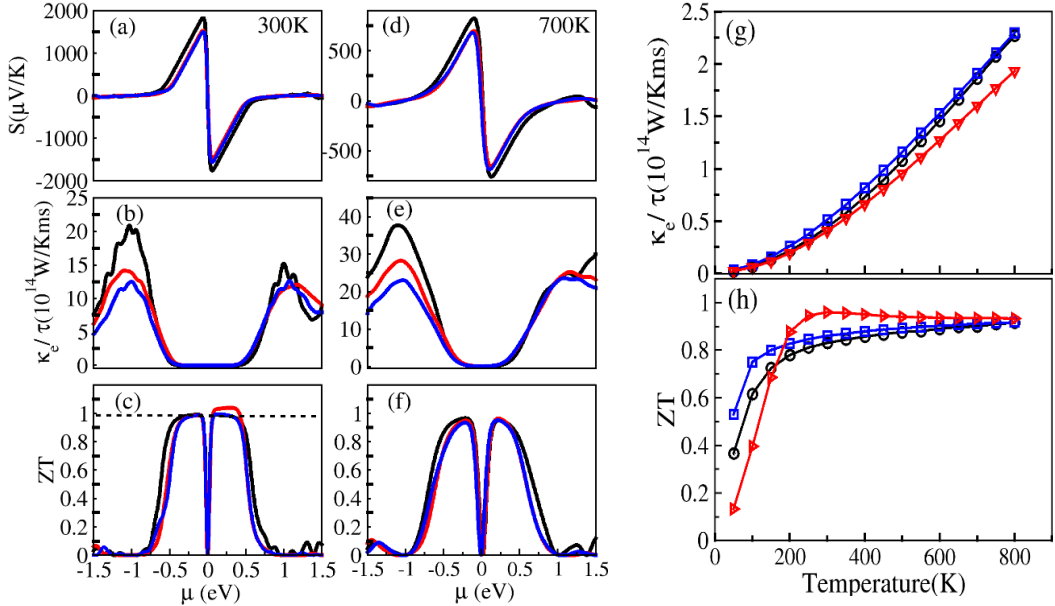

**FIGURE 3.** Transport properties of TiCoSb (black colour), Ti$_{0.5}$Zr$_{0.25}$Hf$_{0.25}$CoPb$_{0.5}$Te$_{0.5}$ (red colour), and Ti$_{0.5}$Zr$_{0.5}$CoPb$_{0.5}$Te$_{0.5}$ (blue colour) as a function of chemical potential at 300K and 700K (left panel) and as a function of temperature (right panel). The calculated transport properties i.e. electrical conductivity, Seebeck coefficient and dimensionless thermoelectric figure of merit were plotted as a function of chemical potential in the range of -1.5 eV to 1.5 eV at 300K and 700K

(see the Fig 3 (a-f)). The Seebeck coefficient of these alloys are almost double at room temperature compared with that at 700K. Dimensionless thermoelectric figure of merit *ZT* is calculated from the Seebeck coefficient (S) electrical conductivity ($\sigma/\tau$), and the electronic part of thermal conductivity ($\kappa_e/\tau$). The electronic thermal conductivity of substituted alloys are smaller than that of the parent alloy TiCoSb (Fig 3 (b, e) ) indicating that by multinary substitution one can decrease the electronic thermal conductivity and hence increase the *ZT* which is required for the higher efficiency thermoelectric materials. Our calculated *ZT* value at room temperature as a function of chemical potential show that the alloy $Ti_{0.5}Zr_{0.25}Hf_{0.25}CoPb_{0.5}Te_{0.5}$ with electron doping (n-type condition) will have high *ZT* of 1.05 which is higher than that of parent compound TiCoSb (~0.98) which indicate that $Ti_{0.5}Zr_{0.25}Hf_{0.25}CoPb_{0.5}Te_{0.5}$ is good candidate for thermoelectric applications at low temperature (see Fig 3 (c, f)). In addition, we have also calculated the electronic thermal conductivity and *ZT* value as a function of temperature as shown in Fig 3 (g, h). The electronic thermal conductivity of n-type $Ti_{0.5}Zr_{0.25}Hf_{0.25}CoPb_{0.5}Te_{0.5}$ is lower than that in other two systems considered for the present study due to the fact that the substitution of (Zr,Hf) and (Pb,Te) introduces additional phonon scattering center into the system. It may be noted from the Fig 3 (h) that *ZT* value is high at 300K and it stay almost constant with increase of temperature.

## CONCLUSION

In conclusion,we have used the VASP code for structural optimizations, WIEN2k for electronic structure calculations and BoltzTraP code for transport calculations to study the substitution effect on $Ti_{0.5}Zr_{0.5}CoPb_{0.5}Te_{0.5}$ $Ti_{0.5}Zr_{0.25}Hf_{0.25}CoPb_{0.5}Te_{0.5}$. We have obtained the electronic structure for TiCoSb and the multinary substituted TiCoSb from WIEN2k calculations and using this the thermoelectric related electrical transport properties including Seebeck coefficients, electrical conductivities, thermoelectric figure of merits, and their dependence on the Fermi levels are calculated by combining the Boltzmann transport theory (BoltzTraP code). The band structures analysis shows that Zr, Hf substituted at Ti site and Pb, Te substituted at Sb site bring the semiconducting behavior having direct band gap feature with VBM and CBM are at $\Gamma$ point. The band gap values of $Ti_{0.5}Zr_{0.5}CoPb_{0.5}Te_{0.5}$ and $Ti_{0.5}Zr_{0.25}Hf_{0.25}CoPb_{0.5}Te_{0.5}$ are 0.92 eV and 0.93 eV which are lower than that of parent TiCoSb (~1.04 eV). From our calculations we suggest that the *ZT* value can be increased in optimally electron doped $Ti_{0.5}Zr_{0.25}Hf_{0.25}CoPb_{0.5}Te_{0.5}$ with the value of 1.05 at room temperature which is higher than that of parent TiCoSb.

## ACKNOWLEDGEMENTS

The authors are grateful to the Department of Science and Technology, India for the funding support via Grant No. SR/NM/NS-1123/2013 and the Research Council of Norway for computing time on the Norwegian supercomputer facilities